Parameter Recovery with Marginal Maximum Likelihood and Markov Chain Monte Carlo

Estimation for the Generalized Partial Credit Model


*Yong Luo*

*National Center for Assessment in Saudi Arabia*



**Abstract**

The generalized partial credit model (GPCM) is a popular polytomous IRT model that has been widely used in large-scale educational surveys and health care services. Same as other IRT models, GPCM can be estimated via marginal maximum likelihood estimation (MMLE) and Markov chain Monte Carlo (MCMC) methods. While studies comparing MMLE and MCMC as estimation methods for other polytomous IRT models such as the nominal response model and the graded response model exist in literature, no studies have compared how well MMLE and MCMC recover the model parameters of GPCM. In the current study, a comprehensive simulation study was conducted to compare parameter recovery of GPCM via MMLE and MCMC. The manipulated factors included latent distribution, sample size, and test length, and parameter recovery was evaluated with bias and root mean square error. It was found that there were no statistically significant differences in recovery of the item location and ability parameters between MMLE and MCMC; for the item discrimination parameter, MCMC had less bias in parameter recovery than MMLE under both normal and uniform latent distributions, and MMLE outperformed MCMC with less bias in parameter recovery under skewed latent






distributions. A real dataset from a high-stakes test was used to demonstrate the estimation of GPCM with MMLE and MCMC.

## Introduction

Approximately three decades ago, Lord (1986) surveyed the estimation methods for item response theory (IRT; 1980) and concluded that the three main approaches were joint maximum likelihood estimation (JMLE; Birnbaum, 1968), marginal maximum likelihood estimation (MMLE; Bock & Aitkin, 1981; Bock & Lieberman, 1970), and Bayesian estimation. More than thirty years later, MMLE is now the dominant estimation method for IRT, and Bayesian estimation has been becoming increasingly popular in the past two decades with the introduction of Markov chain Monte Carlo (MCMC) method into IRT (Albert, 1992; Patz & Junker, 1999a, 1999b), while the use of JMLE for IRT models other than the Rasch model has been gradually phased out due to its inherent estimation bias (cf., Little & Rubin, 1983; Neyman & Scott, 1948).

IRT models are statistical models whose utility depends on the accurate estimation of model parameters, and due to the coexistence of MMLE and MCMC, a natural question facing researchers and practitioners is which estimation method should be employed for their chosen IRT model. To address this question, methodologists have conducted numerous comparison studies to provide concrete advice regarding which estimation method works better under what conditions for a particular IRT model (e.g., Baker, 1998; Jiao, Wang, & He, 2013). However,





one important IRT model that has been widely applied in large-scale educational surveys such as

National Assessment of Educational Progress (NAEP) and health care services such as Patient

Report Outcome Measures (PROM), namely the generalized partial credit model (GPCM;

Muraki, 1992), has remained absent in such comparison studies.

DeMars (2003) conducted the only comprehensive simulation study known to us

regarding parameter recovery of GPCM: she compared how well PARSCALE (Muraki & Bock,

1997) and MULTILOG (Thissen, 1991), two IRT software programs that implements MMLE,

recovered model parameters of the graded response model (GRM; Samejima, 1969) and GPCM.

The manipulated factors in her study include latent distribution (normal, uniform, skewed) and

sample size (250, 500), and 100 replications were implemented within each of the six conditions.

The findings in her study indicate that PARSCALE and MULTILOG produce very similar

estimates for both GRM and GPCM, and she recommended that either software program could

be used. As informative as her study is, however, it does not provide any information regarding

whether MCMC or MMLE should be used for GPCM estimation and, despite the publication of

some influential tutorial articles on how to estimate GPCM with MCMC method implemented in

BUGS (Curtis, 2010; Li & Baser, 2012), no studies comparing the difference between MMLE

and MCMC on GPCM estimation have been conducted in the psychometric literature.

One possible reason for the lack of comparison study between MMLE and MCMC for

the estimation of GPCM is the long computation time required for estimation of complex IRT

models with traditional MCMC methods such as Gibbs sampler. Conventional MCMC methods

usually require a huge number of iterations in each chain for model convergence and





consequently, it takes excessively long time for MCMC estimation, especially in simulation studies where hundreds or thousands of datasets need to be estimated. For example, Wollack, Bolt, Cohen, and Lee (2002) set the number of iterations within each chain to be 11,000 for the estimation of the nominal response model (NRM; Bock, 1972) and, even though their simulation study was of relatively small scale with six simulation conditions (50 replications in each), it took more than two months for 16 computers to run the Gibbs sampler. Although today's computing power is considerably greater than more than a decade ago, a comprehensive simulation study involving the use of traditional MCMC methods still requires an enormous amount of time, which can be practically infeasible for many methodologists.

Hamiltonian Monte Carlo (HMC; Neal, 2011) is a more recent MCMC algorithm that avoids "the random walk behavior" (Gelman, Carlin, Stern, & Rubin, 2014, p. 300) typically found in the traditional MCMC methods. It does so by introducing for each model parameter a momentum variable, which uses the posterior density based on the current drawn parameter value to determine how HMC algorithm efficiently explores the posterior distribution[1]. Hoffman and Gelman (2014) further refined HMC in their proposed No-U-Turn Sampler (NUTS), which is implemented in Stan (Carpenter et al., 2017), an emerging Bayesian software program that is gaining momentum in the psychometric community and has been used in simulation studies (e.g., Bainter, 2017; Luo, 2018a, 2018b; Luo & Al-Harbi, 2017; Revuelta & Ximénez, 2017). One potential advantage that Stan enjoys over other Bayesian software programs implementing the traditional MCMC methods is its computational efficiency: the Stan User Manual claims that "Stan might work fine with 1000 iterations with an example where BUGS would require 100, 000 for good mixing" (Stan Development Team, 2016, p. 541). Luo and Jiao (2018) illustrated in





their tutorial paper that for complex IRT models such as the multilevel three-parameter-logistic (3PL) IRT model (c.f., Fox & Glas, 2001), Stan required as few as 1,000 iterations to achieve model convergence.

In the current study, we fill the gap in the psychometric literature by conducting a comprehensive simulation study to investigate how well MCMC and MML estimation methods recovery the item and ability parameters of the GPCM. The remainder of the current article is organized as follows. First, we show that GPCM is a special case of NRM and hence can be estimated as a multinomial logistic model in Mplus (Muthén & Muthén, 1998-2012), a popular latent variable modeling software program that has been increasingly used for estimation of complex IRT models (e.g, Finch, 2010; Luo, 2018c). Second, we review several simulation studies in the IRT context that compare MMLE and MCMC estimation methods for other polytomous IRT models. Third, we present a simulation study conducted to compare how well MCMC and MMLE recover the model parameters of GPCM. Fourth, we illustrate with a real dataset the parameter comparison from MCMC and MMLE. Last, we conclude the article with discussions and recommendations regarding GPCM estimation in practice.

GPCM as a Special Case of NRM

GPCM, as the name suggests, is a generalized case of the partial credit model (PCM; Masters, 1982) in the sense that the equality constraint on the item discrimination parameter in PCM is freed to allow item-specific discrimination parameter. What the name does not suggest, however, is that GPCM is a special case of NRM. The mathematical equation for NRM is given as





$$p_{ij}(u_{ij} = k \mid \theta_i, a_{jk}, \gamma_{jk}) = \frac{\exp(a_{jk}\theta_i + \gamma_{jk})}{\sum_{h=1}^{c}\exp(a_{jh}\theta_i + \gamma_{jh})}, \qquad (1)$$

where $u_{ij}$ is the response of examinee $i$ to item $j$, $\theta_i$ is the latent proficiency of examinee $i$, $a_{jk}$ and $\gamma_{jk}$ are the slope and intercept parameters of item $j$ on category $k$, and the other terms remain the same.

The mathematical equation for GPCM is given as

$$p_{ij}(u_{ij} = k \mid \theta_i, a_j, \delta_{jk}) = \frac{\exp[\sum_{h=1}^{k_j} a_j(\theta_i - \delta_{jh})]}{\sum_{c=1}^{m_j}\exp[\sum_{h=1}^{c} a_j(\theta_i - \delta_{jh})]}, \qquad (2)$$

where $\delta_{jh}$ is the transition location parameter between categories $k$ and $k$-$1$ of item $j$ with $\delta_{j1}$ set to zero, $m_j$ is the number of categories of item $j$, and the other terms remain the same. Muraki (1992) showed that NRM becomes GPCM if certain constraints are imposed on the slope parameters and consequently, and Huggins-Manley and Algina (2015) demonstrated that Mplus can effectively estimate (with ML estimator that implements MMLE) GPCM as a special case of NRM, which itself belongs to the multinomial logistic family. They also showed that Mplus provided estimation results that were equivalent to those provided by the mainstream IRT estimation software program IRTPRO (Cai, du Toit, & Thissen, 2011) and the general statistical programming software SAS (for estimation of IRT models in SAS before the development of the dedicated procedure PROC IRT, see Sheu, Cheng, Su, & Wang, 2005). In the current study, Mplus is used to implement the MMLE estimation of GPCM.





Previous Simulation Studies on Estimation Comparison of Polytomous IRT Models

In the psychometric literature there are many studies that compare different estimation methods for IRT models (e.g., Albert, 1992; Baker, 1998; de la Torre, Stark, & Chernyshenko, 2006; Jiao, Wang, & He, 2013; Kieftenbeld & Natesan, 2012; Kim, 2001; Kim & Cohen, 1999; Kuo & Sheng, 2016; Wollack, Bolt, Cohen, & Lee, 2003). In this section, we review three simulation studies that compare MMLE and MCMC estimation for polytomous IRT models.

Wollack et al. (2003) compared the performances of MMLE and a MCMC method (Gibbs sampler) in terms of parameter recovery of NRM. They manipulated the sample size (300, 500) and test length (10, 20, 30), and implemented 50 replications within each of the six simulation condition. They found that despite the enormous difference in computation time between these two methods, MMLE and MCMC via Gibbs sampling produced similar parameter estimates. They concluded that for scenarios wherein MMLE is not available, MCMC can be a feasible alternative estimation method.

Kieftenbeld and Natesan (2012) conducted a comprehensive simulation study to compare the performances of MMLE and MCMC via Gibbs sampling regarding parameter recovery of GRM. They manipulated latent trait distribution (normal, skewed, uniform), test length (5, 10, 15, 20), and sample size (75, 150, 300, 500, 1000), which resulted in a fully crossed design with 60 simulation conditions. Within each condition, 100 datasets were generated to be estimated by both MMLE and MCMC. They found that for ability parameters, MMLE/EAP and MCMC produced similar estimates regardless of the test length; for item parameters, MMLE and MCMC also produced similar estimates with sample size no smaller than 300. With sample sizes of 75 or





150, MCMC produced better estimates for some item threshold parameters. They concluded that MCMC and MMLE produced very similar parameter estimates in most conditions, and MCMC had slightly better item parameter recovery with small sample size.

Kuo and Sheng (2016) compared several different estimation methods, namely two MMLE (Bock-Aitkin EM algorithm, adaptive quadrature), four fully Bayesian methods (Gibbs sampler, Metropolis-Hastings, Hastings-within-Gibbs, blocked Metropolis), and Metropolis-Hastings Robins-Monroe, regarding the parameter recovery of a two-dimensional GRT model with a simple structure. The manipulated factors in their simulation study include sample size (500, 1,000), test length (20, 40), and correlation between the two dimensions (0.2, 0.5, 0.8), which resulted in a fully crossed design with twelve conditions. Ten replications were conducted within each condition. Kuo and Sheng found that when the inter-dimension correlation was weak, the seven estimation methods produced similar estimates; when the inter-dimension correlation was moderate or strong, Hastings-within-Gibbs produced better estimates of item discrimination and inter-dimension correlation parameters.

## Methods

### Simulation Design

We conducted a Monte Carlo simulation study to compare how well MMLE and MCMC recovered the model parameters in GPCM. Manipulated factors include latent distribution (normal, skewed, uniform), sample size (SS; 500, 1000, 2000), test length (TL; 5, 10, 20), which results in a fully crossed study with 27 simulation conditions. Under each condition, 100 replications were conducted based on a data generation scheme described in the following.





Data Generation

Item discrimination parameters were generated from a lognormal distribution $\ln N(-0.5, 0.2)$, and the four item location parameters within an item were generated from $N(-1.5, 0.5)$, $N(-0.5, 0.5)$, $N(0.5, 0.5)$, and $N(1.5, 0.5)$, respectively. The generated item parameters are listed in Table 1.

Table 1

*Generating Item Parameter Values*

| Item | a | b1 | b2 | b3 | b4 |
|------|-------|--------|--------|--------|-------|
| 1 | 1.476 | -1.726 | -0.145 | -0.849 | 1.765 |
| 2 | 1.202 | -1.285 | 0.248 | 0.868 | 1.433 |
| 3 | 1.390 | -1.109 | -0.099 | -0.257 | 1.196 |
| 4 | 0.880 | -1.855 | -0.105 | 0.526 | 1.271 |
| 5 | 1.047 | -2.198 | 0.274 | 1.038 | 2.126 |
| 6 | 1.256 | -1.059 | -0.542 | 0.716 | 1.858 |
| 7 | 1.090 | -1.326 | -0.351 | 0.669 | 1.305 |
| 8 | 0.996 | -1.895 | -1.475 | 0.288 | 1.392 |
| 9 | 0.985 | -0.707 | -0.949 | 0.369 | 1.296 |
| 10 | 0.983 | -1.793 | -0.567 | 0.517 | 1.571 |
| 11 | 1.150 | -1.972 | -0.198 | 0.092 | 1.169 |
| 12 | 1.291 | -1.503 | -0.648 | 0.863 | 2.453 |
| 13 | 1.530 | -1.447 | -0.623 | 0.900 | 1.557 |
| 14 | 0.906 | -2.284 | -0.201 | 0.903 | 1.623 |
| 15 | 1.213 | -1.385 | -0.486 | 0.632 | 1.224 |
| 16 | 0.803 | -1.494 | -0.859 | 0.923 | 1.546 |
| 17 | 0.773 | -1.009 | -0.518 | -0.438 | 1.309 |
| 18 | 0.933 | -1.140 | -0.310 | 1.691 | 1.721 |
| 19 | 1.408 | -1.459 | -0.471 | 0.736 | 0.832 |
| 20 | 1.044 | -1.709 | -0.454 | -0.320 | 1.149 |

For ability parameters, when the latent distribution was normal, ability parameters were generated from a standard normal distribution $N(0, 1)$; when the latent distribution was uniform, they were generated from a uniform distribution $U(-3, 3)$; when the latent distribution was





skewed (with a skewness of 1.25 and a kurtosis of 1.5), they were first generated from $N(0, 1)$,

then transformed to be skewed using Fleishman's transformation equation (1978). It should be

noted that with the same latent distribution and sample size, the same set of ability parameters

were used for item response data generation. One hundred datasets under each of the 27

simulation conditions were generated with equation 2.

Estimation

For MMLE estimation, Mplus was used to estimate GPCM as a constrained NRM (see,

Huggins-Manley & Algina, 2015). It should be noted that there are other options that can be used

for MMLE estimation of GPCM, among which there are commercial software programs such as

MULTILOG, PARSCALE, IRTPRO (Cai, du Toit, & Thissen, 2011), and flexMIRT (Cai,

2013), and free solutions such as R packages **mirt** (Chalmers, 2012) and **ltm** (Rizopoulos, 2006).

Based on other researchers' findings (e.g., DeMars, 2002; Huggins-Manley & Algina, 2015) and

our own experience, these software programs produce highly similar estimates for polytmous

IRT models such as GPCM, and one's choice is a matter of personal preference and software

availability.

For the MCMC estimation of GPCM, the following priors were used: a standard normal

distribution $N(0,1)$ was used for the latent ability parameter for model identification; a lognormal

distribution $ln(0,1)$ was used for the item discrimination parameter; a normal distribution with an

unknown mean and standard deviation was used for the item location parameter. For the

unknown mean, a normal distribution $N(0,5)$ was assigned as the hyperprior; for the unknown

standard deviation, a half Cauchy distribution *Cauchy(0,5) w*as used as the hyperprior. We





specified Stan to run three parallel chains with each having 600 iterations (based on our experience, Stan requires approximately 200 iterations to achieve model convergence for GPCM), and the parameter estimates were based on the resultant posterior distribution with 900 draws.

Model Convergence Check

For model convergence check of the MCMC estimation, the potential scale reduction factor (PSRF) proposed by Gelman and Rubin (1992) was used. PSRF values close to one are indicative of model convergence, and 1.1 has been recommend as the cutoff value (Gelman, Carlin, Stern, & Rubin, 2014). In the current study, we used 1.05 as the cutoff value: specifically, after the MCMC estimation was finished for a simulated dataset, we checked the PSRF values to make sure they were all below 1.05; if the PSRF values for some parameters were greater than 1.05, the same dataset would be re-estimated until all PSRF values were smaller than 1.05.

Outcome Variable

Bias and root mean square error (RMSE) regarding parameter estimation were used to compare parameter recovery with MMLE and MCMC. Bias and RMSE are defined as

$$Bias(\hat{\gamma}) = \frac{\sum_1^R(\hat{\gamma}_r - \gamma)}{R}, \tag{3}$$

and
$$RMSE(\hat{\gamma}) = \sqrt{\frac{\sum_1^R(\hat{\gamma}_r - \gamma)^2}{R}}, \tag{4}$$





where $\gamma$ is the true model parameter, $\hat{\gamma}_r$ is the estimated model parameter for the $r$th replication, $\bar{\hat{\gamma}}$ is the mean of model parameter estimates across replications, and $R$ is the number of replications.

In addition, univariate three-way analyses of variance (ANOVA) with both bias and RMSE as dependent variables were conducted to evaluate whether any observed differences across simulation conditions are statistically significant. For evaluation of item parameter recovery, the three factors in ANOVA are estimation method, latent distribution, and sample size; for evaluation of ability parameter recovery, the three factors in ANOVA are estimation method, latent distribution, and test length. Further, Cohen's f was used as a measure of effect size to determine the magnitude of statistically significant differences.

## Results

### Item Location Parameter Recovery

Selected descriptive statistics of bias and RMSE in item location parameter recovery under different latent distributions are listed in Tables 2-4. As can be seen, the average estimation bias for both MMLE and MCMC is close to zero regardless of the latent distribution and sample size; the average RMSE increases with the decrease of sample size. The results of a three-way ANOVA with Bias as the dependent variable indicated that neither the main effects nor the interaction effects were statistically significant regardless of the sample size. In the following we focus on the RMSE in item location parameter estimation with different test lengths.





Table 2

*Bias and RMSE in Item Parameter Estimation under a Normal Latent Distribution*

| TL | Method | SS | Item Discrimination Parameter | | | | Item Location Parameter | | | |
|---|---|---|---|---|---|---|---|---|---|---|
| | | | Bias | | RMSE | | Bias | | RMSE | |
| | | | Mean | SD | Mean | SD | Mean | SD | Mean | SD |
| 20 | MCMC | 2000 | -0.0010 | 0.0052 | 0.0024 | 0.0010 | 0.0003 | 0.0087 | 0.0058 | 0.0032 |
| | | 1000 | -0.0028 | 0.0059 | 0.0050 | 0.0021 | -0.0023 | 0.0139 | 0.0119 | 0.0061 |
| | | 500 | -0.0106 | 0.0088 | 0.0094 | 0.0036 | 0.0021 | 0.0176 | 0.0237 | 0.0128 |
| | MMLE | 2000 | 0.0056 | 0.0058 | 0.0025 | 0.0010 | -0.0001 | 0.0085 | 0.0057 | 0.0032 |
| | | 1000 | 0.0093 | 0.0067 | 0.0052 | 0.0021 | -0.0023 | 0.0116 | 0.0118 | 0.0063 |
| | | 500 | 0.0122 | 0.0092 | 0.0100 | 0.0038 | 0.0022 | 0.0163 | 0.0239 | 0.0142 |
| 10 | MCMC | 2000 | 0.0017 | 0.0075 | 0.0028 | 0.0008 | -0.0002 | 0.0105 | 0.0053 | 0.0023 |
| | | 1000 | -0.0039 | 0.0123 | 0.0058 | 0.0021 | -0.0026 | 0.0120 | 0.0115 | 0.0059 |
| | | 500 | -0.0156 | 0.0146 | 0.0118 | 0.0052 | 0.0005 | 0.0172 | 0.0234 | 0.0124 |
| | MMLE | 2000 | 0.0080 | 0.0083 | 0.0029 | 0.0009 | 0.0000 | 0.0104 | 0.0053 | 0.0023 |
| | | 1000 | 0.0084 | 0.0115 | 0.0060 | 0.0021 | -0.0013 | 0.0124 | 0.0115 | 0.0064 |
| | | 500 | 0.0068 | 0.0118 | 0.0122 | 0.0056 | 0.0015 | 0.0144 | 0.0233 | 0.0135 |
| 5 | MCMC | 2000 | 0.0043 | 0.0079 | 0.0046 | 0.0022 | 0.0025 | 0.0119 | 0.0059 | 0.0033 |
| | | 1000 | -0.0043 | 0.0125 | 0.0102 | 0.0059 | -0.0008 | 0.0097 | 0.0111 | 0.0048 |
| | | 500 | 0.0025 | 0.0257 | 0.0187 | 0.0077 | -0.0025 | 0.0276 | 0.0244 | 0.0124 |
| | MMLE | 2000 | 0.0109 | 0.0086 | 0.0048 | 0.0023 | 0.0018 | 0.0122 | 0.0059 | 0.0033 |
| | | 1000 | 0.0086 | 0.0058 | 0.0106 | 0.0060 | -0.0018 | 0.0083 | 0.0111 | 0.0051 |
| | | 500 | 0.0271 | 0.0213 | 0.0206 | 0.0085 | -0.0050 | 0.0226 | 0.0241 | 0.0133 |





Table 3

*Bias and RMSE in Item Parameter Estimation under a Skewed Latent Distribution*

| TL | Method | SS | Item Discrimination Parameter | | | | Item Location Parameter | | | |
|---|---|---|---|---|---|---|---|---|---|---|
| | | | Bias | | RMSE | | Bias | | RMSE | |
| | | | Mean | SD | Mean | SD | Mean | SD | Mean | SD |
| 20 | MCMC | 2000 | -0.0324 | 0.0185 | 0.0038 | 0.0022 | 0.0009 | 0.0519 | 0.0090 | 0.0053 |
| | | 1000 | -0.0376 | 0.0218 | 0.0067 | 0.0036 | 0.0003 | 0.0539 | 0.0157 | 0.0077 |
| | | 500 | -0.0475 | 0.0226 | 0.0121 | 0.0062 | -0.0006 | 0.0622 | 0.0292 | 0.0143 |
| | MMLE | 2000 | -0.0265 | 0.0173 | 0.0035 | 0.0019 | 0.0039 | 0.0503 | 0.0088 | 0.0052 |
| | | 1000 | -0.0262 | 0.0193 | 0.0060 | 0.0031 | 0.0041 | 0.0510 | 0.0155 | 0.0078 |
| | | 500 | -0.0257 | 0.0175 | 0.0109 | 0.0053 | 0.0049 | 0.0577 | 0.0290 | 0.0159 |
| 10 | MCMC | 2000 | -0.0441 | 0.0275 | 0.0055 | 0.0037 | 0.0036 | 0.0812 | 0.0125 | 0.0085 |
| | | 1000 | -0.0506 | 0.0307 | 0.0088 | 0.0051 | 0.0030 | 0.0864 | 0.0201 | 0.0125 |
| | | 500 | -0.0579 | 0.0324 | 0.0144 | 0.0066 | 0.0040 | 0.0918 | 0.0328 | 0.0165 |
| | MMLE | 2000 | -0.0382 | 0.0266 | 0.0050 | 0.0034 | 0.0046 | 0.0798 | 0.0123 | 0.0086 |
| | | 1000 | -0.0395 | 0.0297 | 0.0079 | 0.0047 | 0.0043 | 0.0843 | 0.0197 | 0.0135 |
| | | 500 | -0.0362 | 0.0313 | 0.0130 | 0.0056 | 0.0046 | 0.0884 | 0.0322 | 0.0184 |
| 5 | MCMC | 2000 | -0.0500 | 0.0431 | 0.0085 | 0.0068 | 0.0083 | 0.1196 | 0.0196 | 0.0152 |
| | | 1000 | -0.0563 | 0.0536 | 0.0137 | 0.0090 | 0.0037 | 0.1302 | 0.0283 | 0.0210 |
| | | 500 | -0.0588 | 0.0386 | 0.0232 | 0.0127 | 0.0091 | 0.1285 | 0.0422 | 0.0236 |
| | MMLE | 2000 | -0.0443 | 0.0423 | 0.0080 | 0.0064 | 0.0067 | 0.1175 | 0.0190 | 0.0152 |
| | | 1000 | -0.0452 | 0.0507 | 0.0126 | 0.0082 | 0.0015 | 0.1261 | 0.0273 | 0.0212 |
| | | 500 | -0.0364 | 0.0365 | 0.0224 | 0.0127 | 0.0058 | 0.1215 | 0.0407 | 0.0249 |





Table 4

*Bias and RMSE in Item Parameter Estimation under a Uniform Latent Distribution*

| TL | Method | SS | Item Discrimination Parameter | | | | Item Location Parameter | | | |
|----|--------|-----|------|------|------|------|------|------|------|------|
| | | | Bias | | RMSE | | Bias | | RMSE | |
| | | | Mean | SD | Mean | SD | Mean | SD | Mean | SD |
| 20 | MCMC | 2000 | 0.0058 | 0.0083 | 0.0024 | 0.0010 | 0.0012 | 0.0218 | 0.0060 | 0.0028 |
| | | 1000 | 0.0014 | 0.0104 | 0.0048 | 0.0020 | 0.0015 | 0.0239 | 0.0117 | 0.0057 |
| | | 500 | -0.0065 | 0.0144 | 0.0096 | 0.0040 | 0.0003 | 0.0272 | 0.0234 | 0.0111 |
| | MMLE | 2000 | 0.0133 | 0.0088 | 0.0026 | 0.0011 | 0.0012 | 0.0230 | 0.0060 | 0.0028 |
| | | 1000 | 0.0144 | 0.0115 | 0.0052 | 0.0021 | 0.0004 | 0.0244 | 0.0117 | 0.0059 |
| | | 500 | 0.0180 | 0.0110 | 0.0103 | 0.0042 | -0.0003 | 0.0274 | 0.0234 | 0.0122 |
| 10 | MCMC | 2000 | 0.0173 | 0.0105 | 0.0035 | 0.0014 | 0.0020 | 0.0380 | 0.0068 | 0.0027 |
| | | 1000 | 0.0138 | 0.0096 | 0.0057 | 0.0021 | 0.0036 | 0.0408 | 0.0121 | 0.0048 |
| | | 500 | 0.0119 | 0.0143 | 0.0120 | 0.0045 | -0.0009 | 0.0389 | 0.0223 | 0.0104 |
| | MMLE | 2000 | 0.0237 | 0.0122 | 0.0039 | 0.0015 | 0.0020 | 0.0381 | 0.0068 | 0.0027 |
| | | 1000 | 0.0258 | 0.0125 | 0.0063 | 0.0025 | 0.0040 | 0.0402 | 0.0120 | 0.0050 |
| | | 500 | 0.0362 | 0.0129 | 0.0139 | 0.0050 | -0.0007 | 0.0386 | 0.0223 | 0.0118 |
| 5 | MCMC | 2000 | 0.0375 | 0.0120 | 0.0061 | 0.0027 | 0.0046 | 0.0582 | 0.0084 | 0.0036 |
| | | 1000 | 0.0429 | 0.0171 | 0.0116 | 0.0053 | 0.0029 | 0.0556 | 0.0140 | 0.0061 |
| | | 500 | 0.0337 | 0.0132 | 0.0214 | 0.0106 | 0.0048 | 0.0592 | 0.0258 | 0.0118 |
| | MMLE | 2000 | 0.0439 | 0.0146 | 0.0068 | 0.0030 | 0.0032 | 0.0582 | 0.0083 | 0.0036 |
| | | 1000 | 0.0561 | 0.0236 | 0.0136 | 0.0068 | 0.0014 | 0.0554 | 0.0140 | 0.0064 |
| | | 500 | 0.0591 | 0.0281 | 0.0255 | 0.0120 | 0.0024 | 0.0574 | 0.0256 | 0.0131 |





As the test length has three levels, three univariate three-way ANOVA with RMSE as the dependent variable were conducted. When TL=20, both latent distribution ($F$ (2, 1442) = 33.845, $p < 0.001$, *Cohen's f* = 0.217) and sample size ($F$ (2, 1422) = 529.840, $p = 0.003$, *Cohen's f* = 0.863) had statistically significant effects upon RMSE in item location parameter estimation. When TL=10, both latent distribution ($F$ (2, 702) = 51.198, $p < 0.001$, *Cohen's f* = 0.381) and sample size ($F$ (2, 702) = 195.019, $p = 0.006$, *Cohen's f* = 0.745) were statistically significant. When TL=5, both latent distribution ($F$ (2, 342) = 46.926, $p < 0.001$, *Cohen's f* = 0.523) and sample size ($F$ (2, 342) = 62.090, $p < 0.001$, *Cohen's f* = 0.602) were statistically significant.

For the main effect of latent distribution, Tukey *post hoc* tests revealed that regardless of test length, the RMSE in item location parameter estimation with skewed latent distribution was significantly greater than those with normal latent distribution ($p < 0.001$) and uniform latent distribution ($p < 0.001$), and there were no statistically significant differences under normal and uniform latent distributions.

For the main effect of sample size, Tukey *post hoc* tests revealed that regardless of test length, the RMSEs in item location parameter estimation with SS=2,000 was significantly smaller than that with SS=1,000 ($p < 0.001$) and SS=500 ($p < 0.001$), and the RMSE in item location parameter estimation with SS=1,000 was also significantly smaller than that with SS=500 ($p < 0.001$).

### Item Discrimination Parameter Recovery

Selected descriptive statistics of bias and RMSE in item discrimination parameter recovery under different latent distributions are also listed in Tables 2-4. As the test length has





three levels, three univariate three-way ANOVA with bias in item discrimination parameter recovery as the dependent variable were conducted. When TL=20, the results indicated that estimation method ($F(1, 342) = 94.983$, $p < 0.001$, *Cohen's f* $= 0.526$), latent distribution ($F(2, 342) = 311.198$, $p < 0.001$, *Cohen's f* $= 1.348$), sample size ($F(2, 342) = 3.039$, $p = 0.049$, *Cohen's f* $= 0.132$), and the interaction between estimation method and sample size ($F(2, 342) = 11.245$, $p < 0.001$, *Cohen's f* $= 0.257$) were statistically significant. When TL=10, estimation method ($F(1, 162) = 21.581$, $p < 0.001$, *Cohen's f* $= 0.366$) and latent distribution ($F(2, 162) = 176.331$, $p < 0.001$, *Cohen's f* $= 1.475$) were statistically significant. When TL=5, estimation method ($F(1, 72) = 5.284$, $p = 0.024$, *Cohen's f* $= 0.270$) and latent distribution ($F(2, 72) = 78.015$, $p < 0.001$, *Cohen's f* $= 1.471$) were statistically significant.

Tukey *post hoc* tests revealed regardless of test length, the bias in item discrimination parameter estimation under different latent distributions were always statistically significant, with those under skewed latent distribution negatively biased, those under uniform latent distribution positively biased, and those with normal latent distribution virtually unbiased.

To further understand how latent distribution and estimation method affected estimation bias in the item discrimination parameter, a visual presentation of the marginal means of Bias under different conditions are plotted in Figure 1. As can be seen, when the latent distribution is normal, MCMC and MMLE have negative and positive estimation bias, respectively, with MCMC having a smaller absolute value. When the latent distribution was skewed, both MMLE and MCMC produced estimates that were negatively biased, with MCMC having a greater absolute value. When the latent distribution was uniform, both MMLE and MCMC produced positively biased estimates, with MMLE having a greater value.





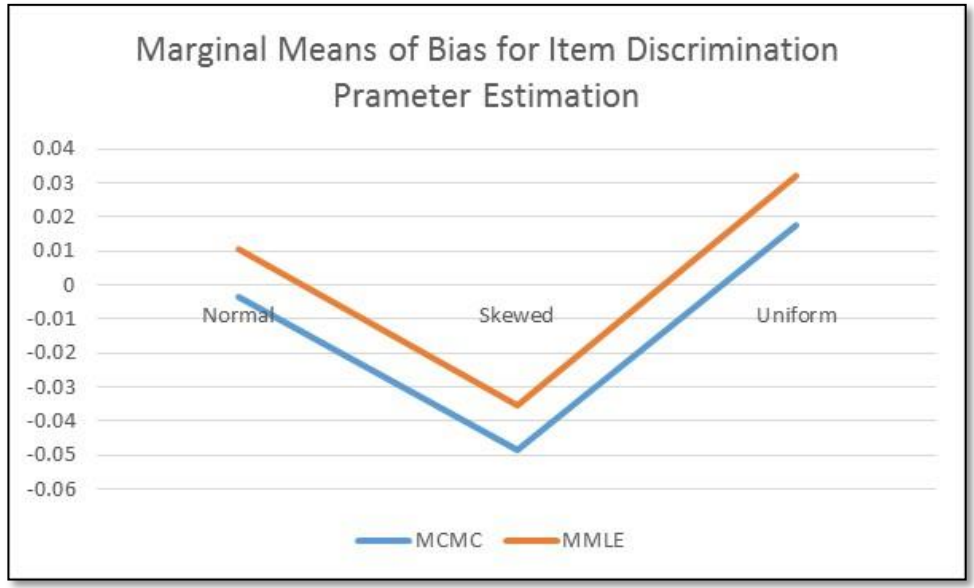

*Figure 1.* Estimation bias in item discrimination parameter

To understand how manipulated factors affected the recovery of item discrimination parameter, three univariate three-way ANOVA with RMSE in item discrimination parameter recovery as the dependent variable were conducted. When TL=20, the results indicated that latent distribution ($F$ (2, 342) = 7.486, $p$ = 0.001, *Cohen's f* = 0.209) and sample size ($F$ (2, 342) = 174.841, $p$ < 0.001, *Cohen's f* = 1.012) were statistically significant. When TL=10, latent distribution ($F$ (2, 162) = 4.973, $p$ = 0.008, *Cohen's f* = 0.248) and sample size ($F$ (2, 162) = 81.504, $p$ < 0.001, *Cohen's f* = 1.004) were statistically significant. When TL=5, only sample size ($F$ (2, 72) = 29.867, $p$ < 0.001, *Cohen's f* = 0.910) were statistically significant.

For the main effect of sample size, Tukey *post hoc* tests revealed that regardless of test length, the RMSE in item discrimination parameter estimation with SS=2,000 was significantly smaller than that with SS=1,000, which was also significantly smaller than that with SS=500.





For the main effect of latent distribution, Tukey *post hoc* tests revealed that when TL=20, the RMSE in item discrimination parameter estimation under a skewed latent distribution was significantly greater than those under both a normal latent distribution ($p = 0.002$) and a uniform latent distribution ($p = 0.003$), and there were no statistically significant differences under a normal and a uniform latent distributions ($p = 0.995$); when TL=10, the RMSE in item discrimination parameter estimation under a skewed latent distribution was significantly greater than that under a normal latent distribution ($p = 0.007$), and there were no statistically significant differences under a normal and a uniform latent distributions ($p = 0.639$) and between a skewed and a uniform latent distributions ($p = 0.080$).

Ability Parameter Recovery

Tables 5 lists selected descriptive statistics of bias in ability parameter estimation under different latent distributions. The results of ANOVA with bias as the dependent variable indicated that neither the main effects nor the interaction effects were statistically significant regardless of the sample size. In the following we mainly discuss RMSE in ability parameter estimation, the descriptive statistics of which are listed in Table 6.





Table 5

*Bias in Ability Parameter Estimation*

| SS | Method | TL | Normal | | Skewed | | Uniform | |
|---|---|---|---|---|---|---|---|---|
| | | | Mean | SD | Mean | SD | Mean | SD |
| 2000 | MCMC | 20 | 0.0003 | 0.0635 | -0.0001 | 0.0854 | 0.0000 | 0.0577 |
| | | 10 | 0.0004 | 0.1135 | -0.0001 | 0.1414 | 0.0002 | 0.1034 |
| | | 5 | -0.0002 | 0.1863 | 0.0000 | 0.2178 | 0.0001 | 0.1702 |
| | MMLE | 20 | 0.0000 | 0.0646 | 0.0024 | 0.0860 | 0.0000 | 0.0593 |
| | | 10 | 0.0006 | 0.1140 | 0.0009 | 0.1417 | 0.0003 | 0.1040 |
| | | 5 | -0.0007 | 0.1862 | -0.0012 | 0.2176 | -0.0011 | 0.1702 |
| 1000 | MCMC | 20 | 0.0002 | 0.0627 | -0.0001 | 0.0901 | 0.0009 | 0.0576 |
| | | 10 | -0.0006 | 0.1123 | -0.0002 | 0.1477 | -0.0002 | 0.1029 |
| | | 5 | -0.0004 | 0.1872 | 0.0000 | 0.2233 | -0.0003 | 0.1690 |
| | MMLE | 20 | 0.0002 | 0.0644 | 0.0029 | 0.0909 | -0.0001 | 0.0597 |
| | | 10 | 0.0007 | 0.1132 | 0.0010 | 0.1480 | 0.0004 | 0.1035 |
| | | 5 | -0.0010 | 0.1867 | -0.0016 | 0.2228 | -0.0014 | 0.1689 |
| 500 | MCMC | 20 | 0.0011 | 0.0637 | -0.0007 | 0.1015 | -0.0001 | 0.0564 |
| | | 10 | -0.0002 | 0.1150 | 0.0005 | 0.1568 | -0.0001 | 0.1033 |
| | | 5 | 0.0002 | 0.1881 | 0.0002 | 0.2315 | 0.0001 | 0.1706 |
| | MMLE | 20 | 0.0008 | 0.0661 | 0.0035 | 0.1029 | -0.0007 | 0.0597 |
| | | 10 | 0.0010 | 0.1156 | 0.0012 | 0.1572 | 0.0004 | 0.1049 |
| | | 5 | -0.0015 | 0.1873 | -0.0022 | 0.2306 | -0.0017 | 0.1705 |





Table 6

RMSE in Ability Parameter Estimation

| SS | Method | TL | Normal | | Skewed | | Uniform | |
|---|---|---|---|---|---|---|---|---|
| | | | Mean | SD | Mean | SD | Mean | SD |
| 2000 | MCMC | 20 | 0.0551 | 0.0213 | 0.0599 | 0.0620 | 0.0538 | 0.0124 |
| | | 10 | 0.1016 | 0.0464 | 0.1125 | 0.1236 | 0.0985 | 0.0243 |
| | | 5 | 0.1713 | 0.0921 | 0.1909 | 0.2173 | 0.1625 | 0.0420 |
| | MMLE | 20 | 0.0550 | 0.0215 | 0.0598 | 0.0621 | 0.0538 | 0.0125 |
| | | 10 | 0.1015 | 0.0466 | 0.1124 | 0.1235 | 0.0984 | 0.0243 |
| | | 5 | 0.1711 | 0.0920 | 0.1907 | 0.2173 | 0.1623 | 0.0420 |
| 1000 | MCMC | 20 | 0.0549 | 0.0216 | 0.0606 | 0.0679 | 0.0543 | 0.0128 |
| | | 10 | 0.1013 | 0.0442 | 0.1144 | 0.1428 | 0.0973 | 0.0239 |
| | | 5 | 0.1726 | 0.0913 | 0.1942 | 0.2500 | 0.1620 | 0.0413 |
| | MMLE | 20 | 0.0548 | 0.0220 | 0.0605 | 0.0681 | 0.0541 | 0.0129 |
| | | 10 | 0.1012 | 0.0445 | 0.1143 | 0.1428 | 0.0972 | 0.0239 |
| | | 5 | 0.1723 | 0.0908 | 0.1940 | 0.2500 | 0.1619 | 0.0414 |
| 500 | MCMC | 20 | 0.0561 | 0.0213 | 0.0636 | 0.1031 | 0.0549 | 0.0125 |
| | | 10 | 0.1023 | 0.0486 | 0.1186 | 0.1927 | 0.0998 | 0.0249 |
| | | 5 | 0.1737 | 0.0944 | 0.1997 | 0.3166 | 0.1638 | 0.0439 |
| | MMLE | 20 | 0.0560 | 0.0219 | 0.0634 | 0.1038 | 0.0548 | 0.0127 |
| | | 10 | 0.1022 | 0.0486 | 0.1184 | 0.1928 | 0.0997 | 0.0251 |
| | | 5 | 0.1735 | 0.0938 | 0.1996 | 0.3158 | 0.1640 | 0.0442 |





As sample size has three levels, three univariate three-way ANOVA with RMSE in ability parameter recovery as the dependent variable were conducted. When SS=2,000, the results of ANOVA indicated that latent distribution ($F$ (2, 35982) = 94.107, $p < 0.001$, *Cohen's f* = 0.071), test length ($F$ (2, 35982) = 4806.132, $p < 0.001$, *Cohen's f* = 0.517), and their interaction ($F$ (4, 35982) = 15.033, $p < 0.001$, *Cohen's f* = 0.045) were statistically significant.

When SS=1,000, latent distribution ($F$ (2, 17982) = 49.131, $p < 0.001$, *Cohen's f* = 0.071), test length ($F$ (2, 17982) = 1944.183, $p < 0.001$, *Cohen's f* = 0.465), and their interaction ($F$ (2, 17982) = 7.648, $p < 0.001$, *Cohen's f* = 0.045) were statistically significant. When SS=500, latent distribution ($F$ (2, 8982) = 20.502, $p < 0.001$, *Cohen's f* = 0.071), test length ($F$ (2, 8982) = 616.078, $p < 0.001$, *Cohen's f* = 0.371), and their interaction ($F$ (2, 8982) = 2.739, $p = 0.027$, *Cohen's f* = 0.032) were statistically significant.

For the main effect of latent distribution, Tukey *post hoc* tests revealed that when SS=2,000 and SS=1,000, the RMSE in ability parameter estimation under a uniform latent distribution was significantly smaller than that under a normal latent distribution ($p = 0.025$) and a skewed latent distribution ($p < 0.001$), and the RMSE in ability parameter estimation under a normal latent distribution was also significantly smaller than that under a skewed latent distribution ($p < 0.001$). With SS=500, the RMSE in ability parameter estimation under a skewed latent distribution was significantly greater than that under a normal latent distribution ($p < 0.001$) and a uniform latent distribution ($p < 0.001$), and there were no statistically significant differences between a normal and a uniform latent distribution (p = 0.405).





For the main effect of test length, Tukey *post hoc* tests revealed that regardless of the sample size, the RMSE in ability parameter estimation with TL=20 was significantly smaller than that with TL=10 ($p < 0.001$) and TL=5 ($p < 0.001$), and the RMSE in ability parameter estimation with TL=10 was also significantly smaller than that with TL=5 ($p < 0.001$).

## A Real Data Example

A dataset extracted from student responses to a test form of the Verbal Session of the General Aptitude Test (GAT-V) was used to demonstrate the comparison of GPCM parameter estimates from MMLE and MCMC. GAT-V, a high-stakes test used in the middle east for college admission purposes, contains 52 multiple-choice items that are dichotomously scored; we created four polytomous items by aggregating the binary items within each of the four reading comprehension passages in the current test form. As each passage contains three items, all four polytomous items have four possible response categories and the score range is from zero to three. We randomly sampled 1,500 students from those who took the current test form. The subsequent analyses were based on this resultant matrix with a dimension of 1,500 by 4.

Mplus and Stan were used again for the implementation of MMLE and MCMC estimation of GPCM, respectively. For estimation with Stan, the same priors used in the previous simulation section were used here, and three parallel chains were specified to run with each containing 600 iterations. The largest PSRF value was 1.013, indicating that model had converged. Consequently, the parameter estimates were based on a posterior sample of 900 draws.





Table 7

*Item Parameter Estimates Comparison*

| Parameter | MMLE | MCMC | Parameter | MMLE | MCMC |
|-----------|------|------|-----------|------|------|
| *alpha1* | 1.213 | 1.203 | *beta2_2* | -0.123 | -0.124 |
| *alpha2* | 1.06 | 1.035 | *beta2_3* | 0.794 | 0.791 |
| *alpha3* | 1.117 | 1.107 | *beta3_1* | -1.628 | -1.641 |
| *alpha4* | 0.3 | 0.323 | *beta3_2* | -0.793 | -0.807 |
| *beta1_1* | -1.894 | -1.907 | *beta3_3* | 1.735 | 1.745 |
| *beta1_2* | -0.998 | -1.008 | *beta4_1* | 0.344 | 0.315 |
| *beta1_3* | 0.261 | 0.251 | *beta4_2* | 2.828 | 2.705 |
| *beta2_1* | -0.969 | -0.977 | *beta4_3* | 6.188 | 5.681 |

Table 7 lists the item parameter estimates with both MMLE and MCMC. As can be seen, the two sets of estimates are not identical but highly similar. While not listed here due to space limitation, the MMLE-based and the MCMC-based ability estimates are also highly similar: the mean difference is 0.003; the standard deviation of the difference is 0.019; the maximum absolute difference is 0.068. As shown in Figure 2, the correlation between the two sets of ability estimates is almost 1. The red dotted regression line (y=x) in Figure 1 is also indicative of the difference between the MMLE-based and the MCMC-based estimates: as can be seen, most points are either on this regression line or clustering tightly around it, an observation which shows that the two sets of ability estimates are virtually the same.





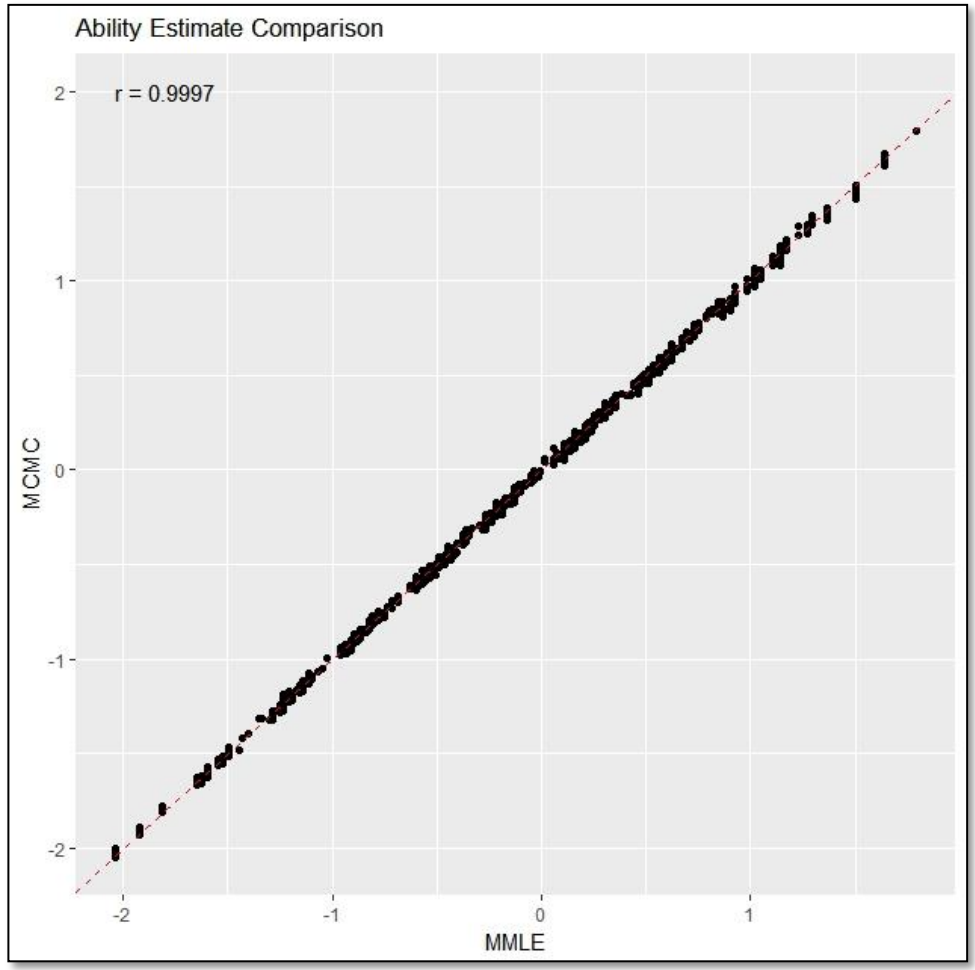

*Figure 2.* Ability estimate comparison based on real data

## Conclusions and Discussions

GPCM is an important and popular polytomous IRT model that has wide applications in

practice. Like other IRT models, GPCM can be estimated with both MMLE and MCMC

methods. However, there are no systematic simulation studies in the psychometric literature that

compare the performances of these two estimation methods regarding parameter recovery of

GPCM. The current study was intended to fill such a gap in literature. Specifically, we conducted





a comprehensive simulation study to investigate whether MMLE and MCMC produce similar parameter estimates. As the implementation of MCMC for complex IRT models is oftentimes extremely time-consuming, we chose Stan, an emerging Bayesian software program which implements the powerful and efficient HMC algorithm and hence runs considerably faster than more traditional Bayesian software programs such as WinBUGS.

The latent distribution was found to have considerable impact upon the parameter recovery quality of GPCM. In terms of estimation bias, the item discrimination parameter recovery was significantly affected by the latent distribution with large effect sizes using Cohen's effect size criteria (1991): when the latent distribution was normal, the estimation bias tended to be positive and close to zero; when the latent distribution was skewed, the estimation bias was negative; when the latent distribution was uniform, the estimation bias was positive.

In terms of estimation RMSE, all GPCM model parameters were significantly affected by the latent distribution, and the parameter recovery quality in terms of RMSE was always worse under a skewed latent distribution than under a normal and a uniform latent distribution. Specifically, the RMSE in the item location parameter estimation under a skewed latent distribution was significantly greater than that under both a normal and a uniform latent distribution regardless of the test length. For item discrimination parameter, the latent distribution had a significant effect upon its estimation RMSE with small effect sizes when the test had twenty or ten items, and the estimation RMSE in item discrimination parameter under a skewed latent distribution was also significantly greater than that under both a normal and a uniform latent distribution. Estimation RMSE in ability parameter was also significantly affected by the latent distribution with small effect sizes regardless of the sample size, and the estimation





RMSE in ability parameter under a skewed latent distribution were again found to be significantly greater than that under a normal and a uniform latent distribution.

Sample size was not found to affect estimation bias in the item discrimination parameter significantly (with the exception when the test had twenty items, where the $p$ value was borderline significant and the effect size was small). For the RMSE in item location and discrimination parameter estimation, however, sample size had significant effects with large effect sizes. This finding is expected, as larger sample sizes provide more information for item parameter estimation and consequently, the resulting item parameter estimates are more stable across samples. Similarly, test length had significant effects with medium to large effect sizes upon the RMSE in the ability parameter estimation, due to the fact that more items lead to more accurate and stable ability estimates.

Regarding the comparison between MMLE and MCMC as estimation methods for GPCM, the simulation results indicated that MMLE and MCMC produced for both the item location parameter and the ability parameter similar estimates that were not statistically significant. For the item discrimination parameter, however, estimation method significantly affected the estimation bias with medium to large effect sizes, the magnitude of which increased with the increase of test length. While MMLE was found to consistently produce item discrimination estimates that were greater in value of estimation bias than their MCMC-based counterparts, the latent distribution affected which estimation method had the smaller absolute value of estimation bias: under both normal and uniform latent distributions, MCMC was the winner as it produced less biased estimates than MMLE; under skewed latent distribution, MMLE outperformed MCMC with its less biased estimates.





We doubt that, however, such differences of estimation bias in the item discrimination parameter between MMLE and MCMC would make much difference in practice, as demonstrated in the real data analysis section. Therefore, the choice of estimation method for GPCM is a matter of software access, computation time, and choice of model selection methods. Regarding software access, while in the current study Mplus, a commercial software program that may be inaccessible to some researchers and practitioners, was used to implement MMLE for GPCM estimation, free alternatives implementing MMLE such as R packages **mirt** and **ltm** can be used. In other words, software programs implementing either MMLE or MCMC are freely available. Regarding computation time, naturally MCMC is considerably slower than MMLE: for one simulated dataset of 2,000 examinees and twenty items in the current study, Mplus took seconds to fit GPCM and Stan took approximately fifteen minutes on a desktop computer with an Intel Xeon E5 processor. The marked time difference notwithstanding, we note that for empirical researchers who are not interested in simulation studies with a large number of conditions, such a difference may not have any practical relevance. If a researcher is interested in model comparison and selection, MCMC may be preferred as it provides more model selection indices than MMLE: MCMC estimates the posterior distribution of each model parameter, which can be used to compute both frequentist-based model selection indices such as Akaike's information criterion (AIC; Akaike, 1973, 1974) and Bayesian information criterion (BIC; Shwarz, 1978) and Bayesian indices such as deviance information criterion (DIC; Spiegelhalter, Best, Carlin, & van der Linde, 2002) and widely available information criterion (WAIC; Watanabe, 2010).





Last but not the least, we would like to reiterate that the HMC sampling algorithm implemented in Stan is what makes a large-scale simulation study like the current one practically feasible: as noted earlier, HMC only required approximately 200 iterations to achieve model convergence for GPCM and consequently, 600 iterations would be adequate for accurate parameter recovery; the more traditional MCMC methods such as Gibbs sampler or Metropolis-Hasting algorithm, however, require thousands of iterations for model convergence (e.g., Kang, Cohen, & Sung, 2009) and hence excessively long computation time. Such long computation time is especially relevant if MCMC methods are used in simulation studies, which usually have a large number of conditions and hundreds of replications within each condition.